\documentclass[12pt]{article}
\usepackage{latexsym}
\usepackage{amsmath,amsfonts}
\usepackage{times}
\allowdisplaybreaks[4]

\catcode`\@=11

\newcount\hour
\newcount\minute
\newtoks\amorpm \hour=\time\divide\hour by 60\minute
=\time{\multiply\hour by 60 \global\advance\minute by-\hour}
\edef\standardtime{{\ifnum\hour<12 \global\amorpm={am}%
        \else\global\amorpm={pm}\advance\hour by-12 \fi
        \ifnum\hour=0 \hour=12 \fi
        \number\hour:\ifnum\minute<10
        0\fi\number\minute\the\amorpm}}
\edef\militarytime{\number\hour:\ifnum\minute<10 0\fi\number\minute}

\def\draftlabel#1{{\@bsphack\if@filesw {\let\thepage\relax
   \xdef\@gtempa{\write\@auxout{\string
      \newlabel{#1}{{\@currentlabel}{\thepage}}}}}\@gtempa
   \if@nobreak \ifvmode\nobreak\fi\fi\fi\@esphack}
        \gdef\@eqnlabel{#1}}
\def\@eqnlabel{}
\def\@vacuum{}
\def\marginnote#1{}
\def\draftmarginnote#1{\marginpar{\raggedright\scriptsize\tt#1}}
\def\draft{
        \pagestyle{plain}
        \overfullrule=2pt
        \oddsidemargin -.5truein
        \def\@oddhead{\sl \phantom{\today\quad\militarytime} \hfil
        \smash{\Large\sl DRAFT} \hfil \today\quad\militarytime}
        \let\@evenhead\@oddhead
        \let\label=\draftlabel
        \let\marginnote=\draftmarginnote
        \def\ps@empty{\let\@mkboth\@gobbletwo
        \def\@oddfoot{\hfil \smash{\Large\sl DRAFT} \hfil}
        \let\@evenfoot\@oddhead}
        \def\@eqnnum{(\theequation)\rlap{\kern\marginparsep\tt\@eqnlabel}%
        \global\let\@eqnlabel\@vacuum}  }

\newcommand{\rf}[1]{(\ref{#1})}
\renewcommand{\theequation}{\thesection.\arabic{equation}}
\renewcommand{\thefootnote}{\fnsymbol{footnote}}
\newcommand{\newsection}{   
\setcounter{equation}{0}\section}

\def\appendix#1{\addtocounter{section}{1}\setcounter{equation}{0}
\renewcommand{\thesection}{\Alph{section}}
\section*{Appendix \thesection\protect\indent \parbox[t]{11.15cm}{#1}}
\addcontentsline{toc}{section}{Appendix \thesection\ \ \ #1}}

\def\be{\begin{equation}}
\def\ee{\end{equation}}
\def\beq{\begin{eqnarray}}
\def\eeq{\end{eqnarray}}

\def\pcheck{\check{p}}

\def\half{{\frac{1}{2}}}
\def\halfsm{\hbox{{\small $\half$}}}

\def\Phik{|\Phi\rangle}
\def\Phibr{\langle\Phi |}

\def\Lambdak{|\Lambda\rangle}

\def\sm#1{{\scriptscriptstyle (#1)}}

\def\Qwt{\widetilde{Q}}

\def\ibf{{\bf i}}
\def\iibf{{\bf ii}}
\def\iiibf{{\bf iii}}
\def\ivbf{{\bf iv}}
\def\vbf{{\bf v}}

\def\sbf{{\bf s}}

\def\cb{\bar{c}}
\def\bb{\bar{b}}

\def\a{{\rm a}}

\def\m{{\rm m}}
\def\min{{\rm min}}
\def\max{{\rm max}}
\def\B{{\scriptscriptstyle {\rm B}}}
\def\FP{{\scriptscriptstyle {\rm FP}}}

\def\tot{{\rm tot}}

\begin{document}


\begin{flushright}
FIAN-TD-2012-08\\
\end{flushright}

\vspace{1cm}

\begin{center}

{\Large \bf BRST-BV approach to cubic interaction vertices

\medskip
for massive and massless higher-spin fields}

\vspace{2.5cm}

R.R. Metsaev\footnote{ E-mail: metsaev@lpi.ru }

\vspace{1cm}

{\it Department of Theoretical Physics, P.N. Lebedev Physical
Institute, \\ Leninsky prospect 53,  Moscow 119991, Russia }

\vspace{3cm}

{\bf Abstract}

\end{center}

Using BRST-BV formulation of relativistic dynamics, we study
arbitrary spin massive and massless fields propagating in flat
space. Generating functions of gauge invariant off-shell cubic
interaction vertices for mixed-symmetry and totally symmetric fields
are obtained. For the case of totally symmetric fields, we derive
restrictions on the allowed values of spins and the number of
derivatives, which provide a classification of cubic interaction
vertices for such fields.  As by product, we present simple
expressions for the Yang-Mills and gravitational interactions of
massive totally symmetric arbitrary spin fields.

\newpage
\renewcommand{\thefootnote}{\arabic{footnote}}
\setcounter{footnote}{0}

\newsection{\large Introduction}

In view of the aesthetic features of higher-spin field theories
\cite{Vasiliev:1990en} these theories have attracted considerable
interest in recent time (for review, see, e.g.,
Refs.\cite{Sagnotti:2011qp,Bekaert:2010hw}). Further progress in
higher-spin field theories requires, among other things, better
understanding of interacting mixed-symmetry field theories. Although
many interesting approaches to the interacting mixed-symmetry fields
are known in the literature analysis of concrete dynamical aspects
of such fields is still a challenging procedure. One of ways to
simplify analysis of mixed-symmetry field dynamics is based on the
use of light-cone gauge approach. In
Refs.\cite{Metsaev:2005ar,Metsaev:2007rn}, using light-cone gauge
approach, we found generating functions of parity invariant cubic
interaction vertices for massive and massless fields of arbitrary
symmetry. Also, we derived restrictions on the allowed values of
spins and the number of derivatives, which provide the complete
classification of cubic interaction vertices for totally symmetric
massless and massive fields. We note however that light-cone gauge
approach, while being powerful to study classical interacting field
theories, becomes cumbersome when considering the quantization and
renormalization of relativistic theories. Therefore, from the
perspective of quantum higher-spin field theories it is desirable to
obtain gauge invariant and manifestly Lorentz invariant off-shell
counterparts of light-cone gauge cubic vertices in
Refs.\cite{Metsaev:2005ar,Metsaev:2007rn}. This is what we do in
this paper.

In order to obtain gauge invariant vertices of massive and massless
fields we use the BRST-BV formulation of field dynamics. The BRST-BV
approach turned out to be successful for the studying manifestly
Lorentz invariant formulation of string theory \cite{Siegel:1984wx}.
In this paper, we demonstrate that it is the BRST-BV method that
provides a possibility for the straightforward
Lorentz-covariantization of light-cone gauge cubic vertices obtained
in Ref.\cite{Metsaev:2005ar}.

\newsection{\large Review of BRST-BV approach to massive and massless
field}

We begin with review the BRST-BV description of free fields
propagating in flat space. In $d$-dimensional Minkowski space, an
arbitrary spin field of the Poincar\'e algebra is labeled by mass
parameter $\m$ and by spin labels $s_1$, \ldots, $s_\nu$. For
massless fields, $\m=0$, $\nu=[\frac{d-2}{2}]$, while for massive
fields, $\m\ne 0$, $\nu=[\frac{d-1}{2}]$. In order to discuss
mixed-symmetry fields it is sufficient to set $\nu
>1 $, while, for the discussion of totally symmetric, fields we can set $\nu=1$.
In what follows, a particular value of $\nu$ does not matter.

{\bf Massless and massive mixed-symmetry fields}. To streamline the
BRST-BV description of mixed-symmetry bosonic fields we use a finite
set of bosonic oscillators $\alpha_n^A$ and fermionic ghost
oscillators $b_n$, $c_n$, $n=1,\ldots,\nu$, for the discussion of
massless fields and a finite set of bosonic oscillators
$\alpha_n^A$, $\zeta_n$, and fermionic ghost oscillators $b_n$,
$c_n$, $n=1,\ldots,\nu$, for the discussion of massive fields (for
notation, see Appendix). Using such oscillators and Grassmann
coordinate $\theta$, we introduce the following ket-vectors to
discuss the mixed-symmetry massless and massive fields:
\beq
\label{man-07052012-01} && |\Phi\rangle \equiv
\Phi(x,\theta,\alpha,b,c)|0\rangle  \hspace{2.3cm} \hbox{massless
field},
\\
\label{man-07052012-02} &&  |\Phi\rangle \equiv
\Phi(x,\theta,\alpha,\zeta,b,c)|0\rangle  \hspace{2cm} \hbox{massive
field}.
\eeq
Ket-vectors \rf{man-07052012-01}, \rf{man-07052012-02} are assumed
to be Grassmann even. Infinite number of ordinary gauge fields
depending of space-time coordinates $x^A$ are obtained by expanding
ket-vectors \rf{man-07052012-01}, \rf{man-07052012-02} into the
Grassmann coordinate $\theta$ and the oscillators $\alpha_n^A$,
$\zeta_n$, $b_n$, $c_n$. In the BRST-BV approach, gauge invariant
action for free massless and massive fields \rf{man-07052012-01},
\rf{man-07052012-02} and corresponding gauge transformations take
the form \cite{Siegel:1984wx}
\beq
\label{man-07052012-03} && S_2 = \half \int d^dx d\theta \Phibr Q_B
\Phik\,, \qquad\quad \Phibr \equiv (\Phik)^\dagger\,,
\\
\label{man-07052012-03n} && \delta \Phik = Q_\B \Lambdak\,,
\eeq
where the BRST operator $Q_\B$ is defined by the relations
\beq
Q_\B & = & \theta (\Box - \m^2) + S^A p^A + \m S + M p_\theta^{}\,,
\qquad p_A\equiv\partial/\partial x^A\,,\qquad
p_\theta\equiv\partial/\partial\theta\,,\qquad
\\
\label{11052012-01} && S^A \equiv \sum_{n=1}^\nu (c_n \bar\alpha_n^A
- \alpha_n^A \cb_n)\,,
\quad
S \equiv \sum_{n=1}^\nu  ( c_n \bar\zeta_n + \zeta_n \cb_n )\,,
\quad M \equiv \sum_{n=1}^\nu c_n \cb_n\,,\qquad
\eeq
$\Box \equiv p^A p^A$. The $\Lambdak$ \rf{man-07052012-03n} is
considered to be Grassmann odd. For massless fields, the $\Lambdak$
depends on $x^A$, $\theta$, $\alpha_n^A$, $b_n$, $c_n$, while, for
massive fields, the $\Lambdak$ depends on $x^A$, $\theta$,
$\alpha_n^A$, $\zeta_n$, $b_n$, $c_n$.

For the ket-vectors \rf{man-07052012-01}, \rf{man-07052012-02} to
describe irreducible fields, some constraints must be imposed on
these ket-vectors.%
\footnote{ Discussion of the constraints and detailed study of
mixed-symmetry massless fields via ket-vector \rf{man-07052012-01}
may be found in Ref.\cite{Alkalaev:2008gi} (see also
Ref.\cite{Sagnotti:2003qa}). Discussion of other approaches to
mixed-symmetry fields may be found in Ref.\cite{Aulakh:1986cb}.}
But to avoid unnecessary complications, we do not impose any
constraints on the ket-vectors. This implies that our ket-vectors
actually describe reducible sets of massless and massive fields.

We now proceed to review the BRST-BV approach to cubic interaction
vertices. To this end we consider the action
\be \label{08052012-01}
S = S_2 + S_3\,,
\ee
where $S_2$ is given in \rf{man-07052012-03}, while cubic
interaction $S_3$  can be presented as
\beq
&& S_3 = \frac{1}{3}\int d1d2d3 \langle\Phi_1| \langle\Phi_2|
\langle\Phi_3|| V_{123}\rangle \,, \qquad dr = d^dx_r d\theta_r\,,
\\
\label{08052012-03} && \qquad |V_{123}\rangle \equiv V_{123}\int
d^dx \delta^{(d)}(x-x_1) \delta^{(d)}(x-x_2)\delta^{(d)}(x-x_3)
|0\rangle_1|0\rangle_2|0\rangle_3\,.\qquad
\eeq
One can make sure that, under gauge transformation given by
\be
\delta \Phik  =   Q_\B \Lambdak - |\Phi \star \Lambda\rangle -
|\Lambda \star \Phi\rangle\,, \qquad |(\Phi\star\Psi)_3\rangle
\equiv \int d1d2 \langle\Phi_1|\langle\Psi_2| |V_{123}\rangle\,,
\ee
action \rf{08052012-01} is invariant (to cubic order in fields)
provided the vertex $|V_{123}\rangle$ satisfies the equations
\be \label{08052012-02}
Q_\B^\tot |V_{123}\rangle =0 \,,\qquad Q_\B^\tot \equiv
\sum_{r=1,2,3} Q_\B^\sm{r}\,.
\ee
Eqs.\rf{08052012-02} tell us that the vertex $|V_{123}\rangle$
should be BRST closed. These equations by themselves do not
determine the vertex $|V_{123}\rangle$ uniquely. Vertices obtained
via field redefinitions take the form $Q_\B^\tot|C_{123}\rangle$ and
such vertices, which we refer to as BRST exact vertices, also
satisfy Eqs.\rf{08052012-02}. Thus all that is required is to find
solutions to Eqs.\rf{08052012-02} which are not BRST exact. It is
such solutions that we discuss in our paper. Solutions for $V_{123}$
\rf{08052012-03} we find can be presented as
\be
V_{123} = V \theta_1\theta_2\theta_3\,,
\ee
where the quantity $V$ depends on i) derivative with respect to
space coordinates $p_r^A$; ii) derivatives with respect to Grassmann
coordinates $p_{\theta_r}^{}$; iii) oscillators $\alpha_n^{\sm{r}
A}$, $\zeta_n^\sm{r}$, $b_n^\sm{r}$, $c_n^\sm{r}$. It is the
quantity $V$ that we refer to as cubic interaction vertex. We now
discuss our solution for the vertex $V$.

\newsection{ \large Cubic vertices for massive and massless fields}

Up to this point our treatment has been applied to vertices for
massive as well as massless fields. Depending on the values of mass
parameters entering cubic vertices, the cubic vertices can be
separated into the following five groups:
\beq
\label{09052012-01} &&  \m_1 = 0\,, \qquad \m_2 = 0,\qquad \m_3 =
0\,;
\\
\label{09052012-02} &&  \m_1 = \m_2 = 0,\qquad \m_3 \ne  0\,;
\\
\label{09052012-03} && \m_1 = \m_2 \equiv \m   \ne 0 ,\qquad \m_3 =
0\,;
\\
\label{09052012-04} && \m_1 \ne 0 ,\qquad \m_2\ne  0,\qquad  \m_1
\ne  \m_2, \qquad \m_3= 0\,;
\\
\label{09052012-04n} && \m_1 \ne 0 ,\qquad \m_2\ne  0,\qquad  \m_3
\ne 0\,.
\eeq
We study cubic vertices having mass parameters as in
\rf{09052012-01}-\rf{09052012-04n} in turn. In what follows we use
the following notation for the operators constructed out of the
oscillators and derivatives:
\beq
&& \hspace{-0.7cm} \pcheck_r^A \equiv p_{r+1}^A - p_{r+2}^A\,,
\qquad \quad \pcheck_{\theta_r}^{} \equiv p_{\theta_{r+1}}^{} -
p_{\theta_{r+2}}^{}\,, \qquad \quad [r \simeq r+3]\,,
\nonumber\\[-9pt]
\label{09052012-10} &&
\\[-9pt]
&& \hspace{-0.7cm}\a_n^{\sm{r} A}\equiv \alpha_n^{\sm{r} A} -
\frac{p_r^A}{\m_r} \zeta_n^\sm{r} \,,\qquad \alpha_{mn}^\sm{rs}
\equiv \alpha_m^{\sm{r} A}\alpha_n^{\sm{s} A}\,,\quad
r,s=1,2,3\,,\qquad m,n,q =1,\ldots,\nu\,.\qquad
\nonumber
\eeq

\subsection{\bf Cubic vertices for three massless fields}

We begin with discussing the parity invariant cubic interaction
vertex for three massless mixed-symmetry fields \rf{09052012-01}.
General solution of Eqs.\rf{08052012-02} takes the form
\beq
\label{09052012-05} V & = & V(L_n^\sm{1}, L_n^\sm{2}, L_n^\sm{3},
Z_{mnq}| Q_{mn}^\sm{11}\,, Q_{mn}^\sm{22}\,, Q_{mn}^\sm{33})\,,
\\
\label{09052012-06} && L_n^\sm{r} = \pcheck_r^A \alpha_n^{\sm{r} A}
+ \pcheck_{\theta_r}^{} c_n^\sm{r}\,,
\\
\label{09052012-07} && Z_{mnq} = Q_{mn}^\sm{12} L_q^\sm{3}  +
Q_{nq}^\sm{23} L_m^\sm{1} + Q_{qm}^\sm{31} L_n^\sm{2}\,,
\\
\label{09052012-08} && Q_{mn}^\sm{rr} = \alpha_{mn}^\sm{rr} +
b_m^\sm{r} c_n^\sm{r} + b_n^\sm{r} c_m^\sm{r} \,,
\\
\label{09052012-09} && \hspace{1.5cm} Q_{mn}^\sm{rr+1} \equiv
\alpha_{mn}^\sm{rr+1} - \half b_m^\sm{r} c_n^\sm{r+1} - \half
b_n^\sm{r+1} c_m^\sm{r}\,,
\eeq
where $V$ in \rf{09052012-05} is arbitrary polynomial of quantities
defined in \rf{09052012-06}-\rf{09052012-08}. The quantities
$\pcheck_r^A$, $\pcheck_{\theta_r}$, $\alpha_{mn}^\sm{rs}$ are
defined in \rf{09052012-10}. Quantities $L_n^\sm{r}$,
$Q_{mn}^\sm{rs}$, and $Z_{mnq}$ are the respective degree 1, 2, and
3 homogeneous polynomials in the oscillators. Henceforth, degree 1,
2, and 3 homogeneous polynomials in the oscillators are referred to
as linear, quadratic, and cubic forms respectively. All forms
appearing in \rf{09052012-05} are BRST closed but not BRST exact.
This implies that solution \rf{09052012-05} cannot be simplified
anymore by using field redefinitions. We note that cubic vertices
depending on the linear forms were discussed in
Ref.\cite{Koh:1986vg}. Comparing solution
\rf{09052012-05}-\rf{09052012-09} with the one obtained in
light-cone gauge (see expressions (5.2),(5.3) in
Ref.\cite{Metsaev:2005ar}), we note that the BRST-BV vertex provides
straightforward Lorentz-covariantization of the light-cone gauge
vertex.

{\bf Vertices for totally symmetric fields}. Because vertex
\rf{09052012-05} has the same structure as the light-cone gauge
vertex in Ref.\cite{Metsaev:2005ar} we can use result in
Ref.\cite{Metsaev:2005ar} to classify vertices of totally symmetric
fields in a rather straightforward way. To consider the totally
symmetric fields it is sufficient to use one sort of oscillators.
Therefore we set $\nu = 1$ in \rf{11052012-01} and ignore
contribution of the oscillators with $n>1$. Also we adopt the
simplified notation for linear forms $L^\sm{r}
\equiv L_1^\sm{r}$ and cubic form $Z\equiv Z_{111}$.%
\footnote{ For totally symmetric massless fields, BRST invariant
linear, quadratic, and cubic forms were discussed in
Refs.\cite{Bengtsson:1987jt,Dempster:2012vw} (see also
Ref.\cite{Buchbinder:2006eq}). Interesting novelty of our
representation for the vertex is that the cubic forms $Z_{mnq}$
\rf{09052012-07} can entirely be presented in terms of linear forms
$L_n^\sm{r}$ \rf{09052012-06} and quadratic forms $Q_{mn}^\sm{rr+1}$
\rf{09052012-09}.}
Now repeating analysis in  Section 5.1 in Ref.\cite{Metsaev:2005ar},
we find a vertex that describes interaction of massless spin
$s^\sm{1}$, $s^\sm{2}$, $s^\sm{3}$ fields,%
\footnote{ For arbitrary $d$, cubic vertices of massless arbitrary
symmetry fields in flat space were found for the first time in the
light-cone gauge in
Refs.\cite{Metsaev:2005ar,Fradkin:1991iy,Metsaev:1993ap} (for $d=4$,
see Ref.\cite{Bengtsson:1983pd}). Manifestly Lorentz invariant
description of cubic vertices for totally symmetric fields was
obtained in Ref.\cite{Manvelyan:2010jr}, while the BRST description
was given in Ref.\cite{Dempster:2012vw} (see also
Ref.\cite{Bengtsson:1987jt}). Manifestly Lorentz invariant on-shell
vertices were discussed in Ref.\cite{Sagnotti:2010at}. In the
framework of BV approach, the discussion of some particular cubic
vertices may be found in Ref.\cite{Bekaert:2005jf}. Interesting use
of the BRST technique for the studying interaction vertices may be
found in Ref.\cite{Polyakov:2010qs}.}
\beq
\label{10052012-01} && \hspace{-1cm} V(s^\sm{1},s^\sm{2},s^\sm{3};k)
= Z^{\frac{1}{2}(\sbf - k)} \prod_{r=1,2,3} (L^\sm{r})^{s^\sm{r} +
\frac{1}{2}(k - \sbf) }\,, \qquad \sbf \equiv \sum_{r=1,2,3}
s^\sm{r}\,, \qquad
\\
\label{10052012-01n} && \sbf - 2s_\min \leq k \leq \sbf \,, \quad
\qquad \sbf - k \qquad \hbox{even integer}\,.
\eeq
From \rf{10052012-01}, we see that there is 1-parameter family of
vertices labeled by non-negative integer $k$ which is the number of
powers of the momenta $p_r^A$, $p_{\theta_r}$ entering the vertices.
Detailed discussion of restrictions \rf{10052012-01n} may be found
in Section 5.1 in Ref.\cite{Metsaev:2005ar}.

\subsection{ Cubic vertices for two massless fields and one massive field}

We now discuss the parity invariant cubic vertices for two massless
and one massive mixed-symmetry fields with mass parameters as in
\rf{09052012-02}. General solution of Eqs.\rf{08052012-02} takes the
form
\be \label{09052012-14}
V = V(L_n^\sm{3}, Q_{mn}^\sm{12}, Q_{mn}^\sm{23}, Q_{mn}^\sm{31}|
Q_{mn}^\sm{11},Q_{mn}^\sm{22}, Q_{mn}^\sm{33})\,,
\ee
where we introduce two representations for linear and quadratic
forms entering cubic vertex \rf{09052012-14}. These representations
are referred to as massive field strength scheme and minimal
derivative scheme. This is to say that, in the massive field
strength scheme and the minimal derivative scheme, linear and
quadratic forms appearing in \rf{09052012-14} take the following
form:
\\
\noindent {\bf Massive field strength scheme}:
\beq
\label{09052012-15} && L_n^\sm{r} = \pcheck_r^A \alpha_n^{\sm{r} A}
+ \pcheck_{\theta_r}^{} c_n^\sm{r}\,, \quad r=1,2\,,\qquad
L_n^\sm{3} = \pcheck_3^A \a_n^{\sm{3} A}\,,
\\
&& Q_{mn}^\sm{12} \equiv \alpha_{mn}^\sm{12} + \frac{1}{2\m_3^2}
L_m^\sm{1} L_n^\sm{2} - \half b_m^\sm{1} c_n^\sm{2} - \half
b_n^\sm{2} c_m^\sm{1} \,,
\\
&& Q_{mn}^\sm{23} = \alpha_m^{\sm{2} A} \a_n^{\sm{3}A} +
\frac{1}{\m_3^2} L_m^\sm{2} p_2^A \a_n^{\sm{3} A}\,,
\\
&& Q_{mn}^\sm{31} = \a_m^{\sm{3}A} \alpha_n^{\sm{1} A}  -
\frac{1}{\m_3^2} p_1^A \a_m^{\sm{3} A} L_n^\sm{1}\,,
\\
&& Q_{mn}^\sm{rr} = \alpha_{mn}^\sm{rr} +  b_m^\sm{r} c_n^\sm{r} +
b_n^\sm{r} c_m^\sm{r} \,, \qquad r =1,2\,,
\\
\label{09052012-20} && Q_{mn}^\sm{33} = \alpha_{mn}^\sm{33} -
\zeta_m^\sm{3} \zeta_n^\sm{3} + b_m^\sm{3} c_n^\sm{3} +  b_n^\sm{3}
c_m^\sm{3} \,.
\eeq
\noindent {\bf Minimal derivative scheme}:
\beq
\label{09052012-21} && L_n^\sm{r} = \pcheck_r^A \alpha_n^{\sm{r} A}
+ \pcheck_{\theta_r}^{} c_n^\sm{r}\,, \qquad r=1,2,3\,,
\\
&& Q_{mn}^\sm{12} \equiv \alpha_{mn}^\sm{12} + \frac{1}{2\m_3^2}
L_m^\sm{1} L_n^\sm{2} - \half b_m^\sm{1} c_n^\sm{2} - \half
b_n^\sm{2} c_m^\sm{1} \,,
\\
&& Q_{mn}^\sm{23} = \alpha_{mn}^\sm{23} -
\frac{\zeta_n^\sm{3}}{2\m_3} L_m^\sm{2} - \frac{1}{2\m_3^2}
L_m^\sm{2} L_n^\sm{3} - \half b_m^\sm{2} c_n^\sm{3} - \half
b_n^\sm{3} c_m^\sm{2}\,,
\\
&& Q_{mn}^\sm{31} = \alpha_{mn}^\sm{31} +
\frac{\zeta_m^\sm{3}}{2\m_3} L_n^\sm{1} - \frac{1}{2\m_3^2}
L_m^\sm{3} L_n^\sm{1}  - \half b_m^\sm{3} c_n^\sm{1} - \half
b_n^\sm{1} c_m^\sm{3} \,,
\\
&& Q_{mn}^\sm{rr} = \alpha_{mn}^\sm{rr} +  b_m^\sm{r} c_n^\sm{r} +
b_n^\sm{r} c_m^\sm{r} \,, \qquad r =1,2\,,
\\
\label{09052012-26} && Q_{mn}^\sm{33} = \alpha_{mn}^\sm{33} -
\zeta_m^\sm{3} \zeta_n^\sm{3} + b_m^\sm{3} c_n^\sm{3} +  b_n^\sm{3}
c_m^\sm{3} \,.
\eeq
Vertex $V$ \rf{09052012-14} is arbitrary polynomial of linear and
quadratic forms appearing in \rf{09052012-14}. The quantities
$\pcheck_r^A$, $\pcheck_{\theta_r}$, $\a_n^{\sm{3}A}$,
$\alpha_{mn}^\sm{rs}$ are defined in \rf{09052012-10}. We make the
following comments.
\\
\ibf) Linear and quadratic forms appearing in \rf{09052012-14} are
BRST closed but not BRST exact. This implies that solution
\rf{09052012-14} cannot be simplified anymore by using field
redefinitions.
\\
\iibf) In massive field strength scheme, the forms $L^\sm{3}$,
$Q^\sm{23}$, $Q^\sm{31}$ are constructed by using the vector
oscillators $\a^{\sm{3}A}$ which are BRST closed but not BRST exact.
Such oscillators streamline the procedure for finding the vertex but
increase number of derivatives entering the vertex.
\\
\iiibf) In the minimal derivative scheme, the linear and quadratic
forms involve minimal number of derivatives. It is not possible to
decrease the number of derivatives by adding BRST closed or BRST
exact expressions to the linear and quadratic forms entering the
minimal derivative scheme.
\\
\ivbf) Linear and quadratic forms in the minimal derivative scheme
differ from the ones in massive field strength scheme by BRST exact
quantities. This implies that vertex in the minimal derivative
scheme is obtained from the one in massive field strength scheme by
using field redefinitions.
\\
\vbf) Comparing minimal derivative solution
\rf{09052012-21}-\rf{09052012-26} with the one obtained in
light-cone gauge (see expressions (6.5)-(6.9) in
Ref.\cite{Metsaev:2005ar}), we note that the BRST-BV vertex provides
straightforward Lorentz-covariantization of the light-cone gauge
vertex.

{\bf Vertices for totally symmetric fields}. To consider totally
symmetric fields we set $\nu = 1$ in \rf{11052012-01} and ignore
contribution of the oscillators with $n>1$. Also we adopt the
simplified notation for linear form $L^\sm{3} \equiv L_1^\sm{3}$ and
quadratic forms $Q^\sm{rr+1}\equiv Q_{11}^\sm{rr+1}$. Using solution
\rf{09052012-21}-\rf{09052012-26} and repeating analysis in Section
6.1 in Ref.\cite{Metsaev:2005ar}, we find vertex that describes
interaction of spin $s^\sm{1}$, $s^\sm{2}$, $s^\sm{3}$ fields with
mass parameters as in \rf{09052012-02},
\beq
\label{10052012-02} && \hspace{-1cm}
V(s^\sm{1},s^\sm{2},s^\sm{3};\tau)  = (L^\sm{3})^\tau
\prod_{r=1,2,3} (Q^\sm{rr+1})^{\sigma^\sm{r+2} } \,, \qquad  \sbf
\equiv \sum_{r=1,2,3} s^\sm{r}\,,
\\
&& \sigma^\sm{r}  = \halfsm (\sbf - \tau) - s^\sm{r} \,,\qquad
r=1,2\,,\qquad \sigma^\sm{3}  = \halfsm (\sbf + \tau) -s^\sm{3} \,,
\\
&& \max(0, s^\sm{3} - s^\sm{1} - s^\sm{2}) \leq \tau \leq s^\sm{3} -
|s^\sm{1} - s^\sm{2}|\,,
\\
&& \sbf - \tau \qquad \hbox{ even integer}\,.
\eeq
From \rf{10052012-02}, we see that there is 1-parameter family of
vertices labeled by non-negative integer $\tau$.

\subsection{ Cubic vertices for one massless and two massive fields
with the same mass values}

General solution of Eqs.\rf{08052012-02} corresponding  to the
parity invariant cubic vertices for one massless and two massive
mixed-symmetry fields with the same mass values \rf{09052012-03} is
given by
\be \label{09052012-27}
V = V(L_n^\sm{1},L_n^\sm{2},L_n^\sm{3}, Q_{mn}^\sm{12},Z_{mnq}|
Q_{mn}^\sm{11},Q_{mn}^\sm{22}, Q_{mn}^\sm{33})\,,
\ee
where  linear, quadratic, and cubic forms appearing in
\rf{09052012-27} take the following form in the massive field
strength scheme and the minimal derivative scheme.
\\
\noindent {\bf Massive field strength scheme}:
\beq
\label{09052012-28} && L_n^\sm{r} = p_3^A \a_n^{\sm{r} A}\,,\quad
r=1,2\,, \qquad L_n^\sm{3} = \pcheck_3^A \alpha_n^{\sm{3}A}
+\pcheck_{\theta_3}^{} c_n^\sm{3}\,,
\\
\label{09052012-29} && Q_{mn}^\sm{12} = \a_m^{\sm{1}A} \a_n^{\sm{2}
A}\,,
\\
\label{09052012-30} && Z_{mnq} = L_m^\sm{1} \a_n^{\sm{2}A}
\alpha_q^{\sm{3}A} - L_n^\sm{2} \a_m^{\sm{1}A} \alpha_q^{\sm{3}A}\,,
\\
&& Q_{mn}^\sm{rr} = \alpha_{mn}^\sm{rr} - \zeta_m^\sm{r}
\zeta_n^\sm{r} +  b_m^\sm{r} c_n^\sm{r} +   b_n^\sm{r} c_m^\sm{r}
\,, \qquad r =1,2\,,
\\
&& Q_{mn}^\sm{33} = \alpha_{mn}^\sm{33}  + b_m^\sm{3} c_n^\sm{3} +
b_n^\sm{3} c_m^\sm{3} \,.
\eeq
\noindent {\bf Minimal derivative scheme}:
\beq
\label{09052012-33} && L_n^\sm{1} = \pcheck_1^A \alpha_n^{\sm{1} A}
+ \m \zeta_n^\sm{1} + \pcheck_{\theta_1}^{} c_n^\sm{1}\,,
\\
&& L_n^\sm{2} = \pcheck_2^A \alpha_n^{\sm{2} A} - \m \zeta_n^\sm{2}
+ \pcheck_{\theta_2}^{} c_n^\sm{2}\,,
\\
&& L_n^\sm{3} = \pcheck_3^A \alpha_n^{\sm{3}A}
+\pcheck_{\theta_3}^{} c_n^\sm{3}\,,
\\
&& Q_{mn}^\sm{12} = \alpha_{mn}^\sm{12} + \frac{\zeta_m^\sm{1}}{2\m}
L_n^\sm{2} - \frac{\zeta_n^\sm{2}}{2\m} L_m^\sm{1} + \zeta_m^\sm{1}
\zeta_n^\sm{2} -\half b_m^\sm{1} c_n^\sm{2} - \half b_n^\sm{2}
c_m^\sm{1} \,, \qquad
\\
&& Z_{mnq} = L_m^\sm{1} \Qwt_{nq}^\sm{23} + L_n^\sm{2}
\Qwt_{qm}^\sm{31} +  L_q^\sm{3} \Qwt_{mn}^\sm{12}\,,
\\
&& \hspace{1.5cm} \Qwt_{mn}^\sm{12} = \alpha_{mn}^\sm{12} +
\zeta_m^\sm{1} \zeta_n^\sm{2} -\half b_m^\sm{1} c_n^\sm{2} -\half
b_n^\sm{2} c_m^\sm{1} \,,
\\
&&  \hspace{1.5cm}  \Qwt_{mn}^\sm{rr+1} = \alpha_{mn}^\sm{rr+1}
-\half b_m^\sm{r} c_n^\sm{r+1} -\half   b_n^\sm{r+1} c_m^\sm{r}
\,,\qquad r=2,3;
\\
&& Q_{mn}^\sm{rr} = \alpha_{mn}^\sm{rr} - \zeta_m^\sm{r}
\zeta_n^\sm{r} +  b_m^\sm{r} c_n^\sm{r} +   b_n^\sm{r} c_m^\sm{r}
\,, \qquad r =1,2\,,
\\
\label{09052012-41} && Q_{mn}^\sm{33} = \alpha_{mn}^\sm{33}  +
b_m^\sm{3} c_n^\sm{3} + b_n^\sm{3} c_m^\sm{3} \,.
\eeq
Vertex $V$ \rf{09052012-27} is arbitrary polynomial of linear,
quadratic, and cubic forms appearing in \rf{09052012-27}. The
quantities $\pcheck_r^A$, $\pcheck_{\theta_r}$, $\a_n^{\sm{r}A}$,
$\alpha_{mn}^\sm{rs}$ are defined in \rf{09052012-10}. We make the
following comments.
\\
\ibf) Linear, quadratic, and cubic forms appearing in
\rf{09052012-27} are BRST closed but not BRST exact.
\\
\iibf) In massive field strength scheme, the forms $L^\sm{1}$,
$L^\sm{2}$, $Q^\sm{12}$, $Z$ are constructed by using the vector
oscillators $\a^{\sm{1}A}$, $\a^{\sm{2}A}$. These oscillators are
BRST closed but not BRST exact.
\\
\iiibf) Linear, quadratic, and cubic forms in the minimal derivative
scheme differ from the ones in massive field strength scheme by BRST
closed and BRST exact quantities. This implies that vertex in the
minimal derivative scheme is obtained from the one in massive field
strength scheme by using field redefinitions and change of vertices
basis.
\\
\ivbf) Cubic form $Z_{mnq}$ \rf{09052012-30} can be rewritten in
terms of field strength for massless field,
\be
Z_{mnq} = F_q^{\sm{3}AB} \a_m^{\sm{1}A} \a_n^{\sm{2}B}\,, \qquad
F_q^{\sm{3}AB} \equiv p_3^A \alpha_q^{\sm{3} B} - p_3^B
\alpha_q^{\sm{3} A}\,.
\ee
\vbf) Comparing minimal derivative solution
\rf{09052012-33}-\rf{09052012-41} with the one obtained in
light-cone gauge (see expressions (6.23)-(6.28) in
Ref.\cite{Metsaev:2005ar}), we note that the BRST-BV vertex provides
straightforward Lorentz-covariantization of the light-cone gauge
vertex.

{\bf Vertices for totally symmetric fields}. To consider totally
symmetric fields we set $\nu = 1$ in \rf{11052012-01} and ignore
contribution of the oscillators with $n>1$. Also we adopt the
simplified notation for linear forms $L^\sm{r} \equiv L_1^\sm{r}$,
quadratic form $Q^\sm{12}\equiv Q_{11}^\sm{12}$, and cubic form $Z
\equiv Z_{111}$. Using solution \rf{09052012-33}-\rf{09052012-41}
and repeating analysis in Section 6.2 in Ref.\cite{Metsaev:2005ar},
we find vertex that describes interaction of spin $s^\sm{1}$,
$s^\sm{2}$, $s^\sm{3}$ fields with mass parameters as in
\rf{09052012-03},
\beq
\label{10052012-03} && \hspace{-1cm}
V(s^\sm{1},s^\sm{2},s^\sm{3}\,;\,k_\min,k_\max) = (Q^\sm{12})^\sigma
Z^\lambda \prod_{r=1,2,3} (L^\sm{r})^{\tau^\sm{r}}\,,
\\
&& \tau^\sm{1}   = k_\max - k_\min - s^\sm{2}\,, \qquad  \tau^\sm{2}
= k_\max - k_\min - s^\sm{1} \,, \qquad \tau^\sm{3}  =
k_\min\,,\qquad
\\
&& \sigma =  \sbf - 2s^\sm{3}  - k_\max + 2 k_\min\,, \qquad \lambda
= s^\sm{3}  - k_\min\,,\qquad \sbf \equiv \sum_{r=1,2,3} s^\sm{r}\,,
\\
&& k_\min  + \max(s^\sm{1},s^\sm{2}) \leq k_\max \leq \sbf-
2s^\sm{3} + 2 k_\min\,,
\qquad
0 \leq k_\min \leq s^\sm{3}\,.
\eeq
From \rf{10052012-03}, we see that there is 2-parameter family of
vertices labeled by non-negative integers $k_\min$ and $k_\max$
which are the respective the minimal and maximal numbers of powers
of momenta $p_r^A$, $p_{\theta_r}$. Vertex $V(0,0,1;1,1)=L^\sm{3}$
describes Yang-Mills interaction of massive spin-0 field, while
vertex $V(s,s,1;0,s)=(Q^\sm{12})^{s-1}Z$ is a candidate for
Yang-Mills interaction of massive spin-$1$, $s\geq 1$, field.
Vertices $V(0,0,2;2,2)=(L^\sm{3})^2$ and $V(1,1,2; 1, 2) = L^\sm{3}Z
$ describe gravitational interaction of the respective spin-0 and
spin-1 massive fields, while vertex $V(s,s,2; 0,s) =
(Q^\sm{12})^{s-2}Z^2$ is a candidate for gravitational interaction
of massive spin-$s$, $s\geq 2$, field. Recent discussion of
electro-magnetic interaction of totally symmetric massive field may
be found in Ref.\cite{Buchbinder:2012iz}. Discussion of some
particular cases of cubic vertices may be found in
Ref.\cite{Metsaev:2006ui}.

\subsection{ Cubic vertices for one massless and two massive fields
with different mass values}

General solution of Eqs.\rf{08052012-02} corresponding  to the
parity invariant cubic vertices for one massless and two massive
mixed-symmetry fields with different mass values \rf{09052012-04} is
given by
\be \label{09052012-42}
V = V(L_n^\sm{1},L_n^\sm{2},
Q_{mn}^\sm{12},Q_{mn}^\sm{23},Q_{mn}^\sm{31}|
Q_{mn}^\sm{11},Q_{mn}^\sm{22}, Q_{mn}^\sm{33})\,,
\ee
where  linear and quadratic forms appearing in \rf{09052012-42} take
the following form in the massive field strength scheme and the
minimal derivative scheme.
\\
\noindent {\bf Massive field strength scheme}:
\beq
\label{09052012-43} && L_n^\sm{r} = p_3^A \a_n^{\sm{r} A}\,,\quad
r=1,2\,, \qquad L_n^\sm{3} = \pcheck_3^A \alpha_n^{\sm{3}A}
+\pcheck_{\theta_3}^{} c_n^\sm{3}\,,
\\
&& Q_{mn}^\sm{12} = \a_m^{\sm{1}A} \a_n^{\sm{2} A}\,,
\\
&& Q_{mn}^\sm{23} = \a_m^{\sm{2}A} \alpha_n^{\sm{3} A} +
\frac{1}{\m_1^2 - \m_2^2} L_m^\sm{2} L_n^\sm{3}\,,
\\
&& Q_{mn}^\sm{31} = \alpha_m^{\sm{3} A} \a_n^{\sm{1}A}  +
\frac{1}{\m_1^2 - \m_2^2} L_m^\sm{3} L_n^\sm{1} \,,
\\
&& Q_{mn}^\sm{rr} = \alpha_{mn}^\sm{rr} - \zeta_m^\sm{r}
\zeta_n^\sm{r} +  b_m^\sm{r} c_n^\sm{r} +   b_n^\sm{r} c_m^\sm{r}
\,, \qquad r =1,2\,,
\\
\label{09052012-48} && Q_{mn}^\sm{33} = \alpha_{mn}^\sm{33}  +
b_m^\sm{3} c_n^\sm{3} + b_n^\sm{3} c_m^\sm{3} \,.
\eeq
\noindent {\bf Minimal derivative scheme}:
\beq
\label{09052012-49} && L_n^\sm{1} = \pcheck_1^A \alpha_n^{\sm{1}A} +
\frac{\m_2^2}{\m_1}\zeta_n^\sm{1} + \pcheck_{\theta_1}^{} c_n^\sm{1}
\,,
\\
&& L_n^\sm{2} = \pcheck_2^A \alpha_n^{\sm{2}A} -
\frac{\m_1^2}{\m_2}\zeta_n^\sm{2} + \pcheck_{\theta_2}^{}
c_n^\sm{2}\,,
\\
&& L_n^\sm{3} = \pcheck_3^A \alpha_n^{\sm{3}A}
+\pcheck_{\theta_3}^{} c_n^\sm{3}\,,
\\
&& Q_{mn}^\sm{12} = \alpha_{mn}^\sm{12} +
\frac{\zeta_m^\sm{1}}{2\m_1} L_n^\sm{2} -
\frac{\zeta_n^\sm{2}}{2\m_2} L_m^\sm{1} + \frac{\m_1^2 +
\m_2^2}{2\m_1\m_2} \zeta_m^\sm{1} \zeta_n^\sm{2}  -\half b_m^\sm{1}
c_n^\sm{2} - \half b_n^\sm{2} c_m^\sm{1} \,, \qquad
\\
&& Q_{mn}^\sm{23} = \alpha_{mn}^\sm{23} +
\frac{\zeta_m^\sm{2}}{2\m_2} L_n^\sm{3} + \frac{1}{2(\m_1^2 -
\m_2^2)} L_m^\sm{2} L_n^\sm{3} -\half b_m^\sm{2} c_n^\sm{3} - \half
b_n^\sm{3} c_m^\sm{2} \,, \qquad
\\
&& Q_{mn}^\sm{31} = \alpha_{mn}^\sm{31} -
\frac{\zeta_n^\sm{1}}{2\m_1} L_m^\sm{3} - \frac{1}{2(\m_1^2 -
\m_2^2)} L_m^\sm{3} L_n^\sm{1}  - \half b_m^\sm{3} c_n^\sm{1} -\half
b_n^\sm{1} c_m^\sm{3}  \,, \qquad
\\
&& Q_{mn}^\sm{rr} = \alpha_{mn}^\sm{rr} - \zeta_m^\sm{r}
\zeta_n^\sm{r} +  b_m^\sm{r} c_n^\sm{r} +   b_n^\sm{r} c_m^\sm{r}
\,, \qquad r =1,2\,,
\\
\label{09052012-56} && Q_{mn}^\sm{33} = \alpha_{mn}^\sm{33}  +
b_m^\sm{3} c_n^\sm{3} + b_n^\sm{3} c_m^\sm{3} \,.
\eeq
Vertex $V$ \rf{09052012-42} is arbitrary polynomial of linear and
quadratic forms appearing in \rf{09052012-42}. The quantities
$\pcheck_r^A$, $\pcheck_{\theta_r}$, $\a_n^{\sm{r}A}$,
$\alpha_{mn}^\sm{rs}$ are defined in \rf{09052012-10}. We make the
following comments.
\\
\ibf) Linear and quadratic forms appearing in \rf{09052012-42} are
BRST closed but not BRST exact.
\\
\iibf) In massive field strength scheme, the forms $L^\sm{1}$,
$L^\sm{2}$, $Q^\sm{12}$,  $Q^\sm{23}$,  $Q^\sm{31}$, are constructed
by using the vector oscillators $\a^{\sm{1}A}$, $\a^{\sm{2}A}$.
These oscillators are BRST closed but not BRST exact.
\\
\iiibf) Linear and quadratic forms in the minimal derivative scheme
differ from the ones in massive field strength scheme by BRST exact
quantities.
\\
\ivbf) Comparing minimal derivative solution
\rf{09052012-49}-\rf{09052012-56} with the one obtained in
light-cone gauge (see expressions (6.57)-(6.63) in
Ref.\cite{Metsaev:2005ar}), we note that the BRST-BV vertex provides
straightforward Lorentz-covariantization of the light-cone gauge
vertex.

{\bf Vertices for totally symmetric fields}. To consider totally
symmetric fields we set $\nu = 1$ in \rf{11052012-01} and ignore
contribution of the oscillators with $n>1$. Also we adopt the
simplified notation for linear forms $L^\sm{1} \equiv L_1^\sm{1}$,
$L^\sm{2} \equiv L_1^\sm{2}$ and quadratic forms $Q^\sm{rr+1}\equiv
Q_{11}^\sm{rr+1}$. Using solution \rf{09052012-49}-\rf{09052012-56}
and repeating analysis in Section 6.3 in Ref.\cite{Metsaev:2005ar},
we find a vertex that describes interaction of spin $s^\sm{1}$,
$s^\sm{2}$, $s^\sm{3}$ fields with mass parameters as in
\rf{09052012-04},
\beq
\label{10052012-04} && \hspace{-1cm}
V(s^\sm{1},s^\sm{2},s^\sm{3}\,;\,\tau^\sm{1} ,\tau^\sm{2}) =
(L^\sm{1})^{\tau^\sm{1}} (L^\sm{2})^{\tau^\sm{2}} \prod_{r=1,2,3}
(Q^\sm{rr+1})^{\sigma^\sm{r+2}}\,,
\\
&& \sigma^\sm{1} = \halfsm (s^\sm{2}  + s^\sm{3} -s^\sm{1}  +
\tau^\sm{1} - \tau^\sm{2})\,,
\\
&& \sigma^\sm{2} = \halfsm (s^\sm{1}  + s^\sm{3} -s^\sm{2} -
\tau^\sm{1} + \tau^\sm{2} )\,,
\\
&& \sigma^\sm{3} = \halfsm (s^\sm{1}  + s^\sm{2} -s^\sm{3} -
\tau^\sm{1} - \tau^\sm{2})\,,
\\
&& |s^\sm{1}  -s^\sm{2}  - \tau^\sm{1}   + \tau^\sm{2} | \leq
s^\sm{3} \leq s^\sm{1} + s^\sm{2} - \tau^\sm{1}   - \tau^\sm{2}  \,,
\\
&& \sbf - \tau^\sm{1}   - \tau^\sm{2}   \qquad \hbox{ even
integer}\,, \qquad \sbf \equiv s^\sm{1}+s^\sm{3}+s^\sm{3}\,.
\eeq
Thus, there is 2-parameter family of vertices \rf{10052012-04}
labeled by non-negative integers $\tau^\sm{1}$, $\tau^\sm{2}$.

\subsection{ Cubic vertices for three massive fields}

General solution of Eqs.\rf{08052012-02} corresponding  to the
parity invariant cubic vertices for three massive mixed-symmetry
fields \rf{09052012-04n} is given by
\be \label{09052012-57}
V = V(L_n^\sm{1},L_n^\sm{2},L_n^\sm{3},
Q_{mn}^\sm{12},Q_{mn}^\sm{23},Q_{mn}^\sm{31}|Q_{mn}^\sm{11},Q_{mn}^\sm{22},
Q_{mn}^\sm{33})\,,
\ee
where  linear and quadratic forms appearing in \rf{09052012-57} take
the following form in the massive field strength scheme and the
minimal derivative scheme.
\\
\noindent {\bf Massive field strength scheme}:
\beq
\label{09052012-58}  && L_n^\sm{r} = \pcheck_r^A \a_n^{\sm{r} A}\,,
\qquad Q_{mn}^\sm{rr+1} = \a_m^{\sm{r}A} \a_n^{\sm{r+1} A}\,,
\\
\label{09052012-59}  && Q_{mn}^\sm{rr} = \alpha_{mn}^\sm{rr} -
\zeta_m^\sm{r} \zeta_n^\sm{r} +  b_m^\sm{r} c_n^\sm{r} + b_n^\sm{r}
c_m^\sm{r} \,.
\eeq
\noindent {\bf Minimal derivative scheme}:
\beq
\label{09052012-60}  && L_n^\sm{r} = \pcheck_r^A \alpha_n^{\sm{r} A}
+ \pcheck_{\theta_r}^{} c_n^\sm{r} + \frac{\m_{r+1}^2 -
\m_{r+2}^2}{\m_r} \zeta_n^\sm{r}\,,
\\
&& Q_{mn}^\sm{rr+1} = \alpha_{mn}^\sm{rr+1} +
\frac{\zeta_m^\sm{r}}{2\m_r} L_n^\sm{r+1} -
\frac{\zeta_n^\sm{r+1}}{2\m_{r+1}} L_m^\sm{r} +
\frac{\zeta_m^\sm{r}\zeta_n^\sm{r+1}}{2\m_r \m_{r+1}} (\m_r^2 +
\m_{r+1}^2 - \m_{r+2}^2)
\nonumber\\
&& \hspace{1cm} - \, \half b_m^\sm{r} c_n^\sm{r+1} - \half
b_n^\sm{r+1} c_n^\sm{r}\,,
\\
\label{09052012-63}  && Q_{mn}^\sm{rr} = \alpha_{mn}^\sm{rr} -
\zeta_m^\sm{r} \zeta_n^\sm{r} +  b_m^\sm{r} c_n^\sm{r} + b_n^\sm{r}
c_m^\sm{r} \,.
\eeq
Vertex $V$ \rf{09052012-57} is arbitrary polynomial of linear and
quadratic forms appearing in \rf{09052012-57}. The quantities
$\pcheck_r^A$, $\pcheck_{\theta_r}$, $\a_n^{\sm{r}A}$,
$\alpha_{mn}^\sm{rs}$ are defined in \rf{09052012-10}. We make the
following comments.
\\
\ibf) Linear and quadratic forms appearing in \rf{09052012-57} are
BRST closed but not BRST exact.
\\
\iibf) In massive field strength scheme, the forms $L^\sm{r}$,
$Q^\sm{rr+1}$ are constructed by using the vector oscillators
$\a^{\sm{r}A}$. These oscillators are BRST closed but not BRST
exact.
\\
\iiibf) Linear and quadratic forms in the minimal derivative scheme
differ from the ones in massive field strength scheme by BRST exact
quantities.
\\
\ivbf) Comparing minimal derivative solution
\rf{09052012-60}-\rf{09052012-63} with the one in light-cone gauge
(see expressions (7.2)-(7.4) in Ref.\cite{Metsaev:2005ar}), we note
that the BRST-BV vertex provides straightforward
Lorentz-covariantization of the light-cone gauge vertex.

{\bf Vertices for totally symmetric fields}. To consider totally
symmetric fields we set $\nu = 1$ in \rf{11052012-01} and ignore
contribution of the oscillators with $n>1$. Also we adopt the
simplified notation for linear forms $L^\sm{r} \equiv L_1^\sm{r}$
and quadratic forms $Q^\sm{rr+1}\equiv Q_{11}^\sm{rr+1}$. Using
solution \rf{09052012-60}-\rf{09052012-63} and repeating analysis in
Section 7.1 in Ref.\cite{Metsaev:2005ar}, we find a vertex that
describes interaction of massive spin $s^\sm{1}$, $s^\sm{2}$,
$s^\sm{3}$ fields,
\beq
\label{10052012-05} && \hspace{-1cm}
V(s^\sm{1},s^\sm{2},s^\sm{3};\tau^\sm{1},\tau^\sm{2},\tau^\sm{3}) =
\prod_{r=1,2,3}
(L^\sm{r})^{\tau^\sm{r}}(Q^\sm{rr+1})^{\sigma^\sm{r+2}}\,,
\\
&& \sigma^\sm{r}  = \halfsm (\sbf + \tau^\sm{r} - \tau^\sm{r+1} -
\tau^\sm{r+2}) - s^\sm{r} \,,
\\
&& s^\sm{3}  - s^\sm{1} - s^\sm{2}  + \tau^\sm{1} + \tau^\sm{2} \leq
\tau^\sm{3} \leq s^\sm{3} - | s^\sm{1} - s^\sm{2} - \tau^\sm{1} +
\tau^\sm{2} |\,,
\\
&& \sbf + \sum_{r=1,2,3} \tau^\sm{r}  \quad \hbox{ even integer};
\qquad
\sbf \equiv \sum_{r=1,2,3} s^\sm{r}\,.
\eeq
Thus, there is 3-parameter family of vertices \rf{10052012-05}
labeled by non-negative integers $\tau^\sm{1}$, $\tau^\sm{2}$,
$\tau^\sm{3}$.

To summarize, using the BRST-BV approach, we obtained
Lorentz-covariant off-shell description of vertices for massive and
massless fields which were obtained in light-cone gauge in
Ref.\cite{Metsaev:2005ar}. We think that the BRST-BV approach we
discussed in this paper might be useful for the studying interaction
vertices of AdS fields. Recent discussion of interaction vertices of
AdS fields may be found in Ref.\cite{Alkalaev:2010af}. Interesting
use of BRST technique in AdS space may be found in
Ref.\cite{Buchbinder:2006ge}.

{\bf Acknowledgments}. This work was supported by the RFBR Grant
No.11-02-00814.

\setcounter{section}{0}\setcounter{subsection}{0}
\appendix{ \large Notation }

Our conventions are as follows. $x^A$ denotes coordinates in
$d$-dimensional flat space-time, while $p_A$ denotes derivatives
with respect to $x^A$, $p_A \equiv
\partial/\partial x^A$. Vector indices of the Lorentz algebra
$so(d-1,1)$ take the values $A,B,C=0,1,\ldots ,d-1$. To simplify our
expressions we drop mostly positive flat metric tensor
$\eta_{AB}^{}$ in scalar products: $X^AY^A \equiv \eta_{AB}X^A Y^B$.
We use the Grassmann coordinate $\theta$, $\theta^2=0$, and the
corresponding derivative $p_\theta=\partial/\partial\theta$,
$\{p_\theta,\theta\}=1$. Integration in $\theta$ is defined as $\int
d\theta \theta=1$. We use a set of the creation bosonic operators
$\alpha_n^A$, $\zeta_n$ and fermionic ghost operators $b_n$, $c_n$
and the respective set of annihilation bosonic operators
$\bar\alpha_n^A$, $\bar\zeta_n$ and fermionic ghost operators
$\cb_n$, $\bb_n$. These
operators are referred to as oscillators in this paper.%
\footnote{ Interesting study and applications of the oscillator
formalism may be fond in Ref.\cite{Bekaert:2006ix}.}
(Anti)commutation relations, the vacuum, and hermitian conjugation
rules are defined as
\beq
&& [\bar\alpha_m^A,\alpha_n^B] = \eta^{AB}\delta_{mn} \,, \quad
[\bar\zeta_m,\zeta_n] = \delta_{mn}\,,\qquad \{\bb_m, c_n\} =
\delta_{mn}\,, \quad \{ \cb_m, b_n\} = \delta_{mn}\,,\qquad
\\
&& \bar\alpha_n^A|0\rangle = 0\,, \qquad \bar\zeta_n|0\rangle = 0\,,
\qquad \bb_n|0\rangle = 0\,, \qquad \cb_n|0\rangle = 0\,,
\\
&& \alpha_n^{A\dagger} = \bar\alpha_n^A\,,\qquad \zeta_n^\dagger =
\bar\zeta_n\,,\qquad b_n^\dagger = \bb_n\,,\qquad c_n^\dagger =
\cb_n\,.\qquad
\eeq
For momenta and coordinates, we use the hermitian conjugation rules
$p^{A\dagger} = - p^A$, $p_\theta^\dagger = p_\theta$, $x^{A\dagger}
= x^A$, $\theta^\dagger = \theta$. Hermitian conjugation rule for
the product of two arbitrary ghost parity operators $A$, $B$ is
defined as $(AB)^\dagger = B^\dagger A^\dagger$. The ghost number
operator is defined as
\be
N_\FP \equiv \theta p_\theta^{} + N_c - N_b\,, \qquad N_b \equiv
\sum_{n=1}^\nu b_n\cb_n\,, \qquad N_c \equiv \sum_{n=1}^\nu
c_n\bb_n\,.
\ee

\end{document}